\documentclass[10pt,journal]{IEEEtran}
\usepackage{setspace}
\usepackage{clrscode}
\usepackage{pict2e}

\usepackage{amsmath}
\usepackage{amssymb}
\usepackage[dvips]{graphicx}
\usepackage{epsfig}
\usepackage{hyperref}

\usepackage[ps2pdf,
bookmarks=false,
bookmarksnumbered=false, 
bookmarksopen=false, 
colorlinks=false]{}

\newcommand{\E}{\mathbb{E}}
\newcommand{\e}{\epsilon}

\newcommand{\Pb}{\overline{P}}

\newcommand{\figsize}{0.48}
\newcommand{\ssnr}{\text{\footnotesize{SNR}}}

\newcommand{\tSNR}{\text{\footnotesize{SNR}}}
\newcommand{\mH}{\mathcal{H}}
\newcommand{\hH}{\hat{\mathcal{H}}}

\newcommand{\y}{\mathbf{y}}
\newcommand{\x}{\mathbf{x}}
\newcommand{\n}{\mathbf{n}}
\newcommand{\s}{\mathbf{s}}
\newcommand{\Q}{\mathbb{Q}}
\newcommand{\CN}{\mathcal{CN}}

\makeatletter

\newcommand{\Rmnum}[1]{\expandafter\@slowromancap\romannumeral #1@}
\makeatother

\begin{document}
\title{Throughput of Cognitive Radio Systems with Finite Blocklength Codes}

\author{
\authorblockN{Gozde Ozcan and M. Cenk Gursoy}
\thanks{Manuscript received November 18, 2012; revised April 4, 2013.}
\thanks{The authors are with the Department of Electrical
Engineering and Computer Science, Syracuse University, Syracuse, NY, 13244
(e-mail: gozcan@syr.edu, mcgursoy@syr.edu).}
\thanks{This work was supported by the National Science Foundation under Grants CNS -- 0834753 and CCF-0917265. The material in this paper was presented in part at the 2012 Conference on Information Sciences and Systems (CISS), Princeton University, Princeton, NJ.
}}
\date{}

\maketitle
\begin{abstract}
In this paper, throughput achieved in cognitive radio channels with finite blocklength codes under buffer limitations is studied. Cognitive users first determine the activity of the primary users' through channel sensing and then initiate data transmission at a power level that depends on the channel sensing decisions. It is assumed that finite blocklength codes are employed in the data transmission phase. Hence, errors can occur in reception and retransmissions can be required. Primary users' activities are modeled as a two-state Markov chain and an eight-state Markov chain is constructed in order to model the cognitive radio channel. Channel state information (CSI) is assumed to be perfectly known by either the secondary receiver only or both the secondary transmitter and receiver. In the absence of CSI at the transmitter, fixed-rate transmission is performed whereas under perfect CSI knowledge, for a given target error probability, the transmitter varies the rate according to the channel conditions. Under these assumptions, throughput in the presence of buffer constraints is determined by characterizing the maximum constant arrival rates that can be supported by the cognitive radio channel while satisfying certain limits on buffer violation probabilities. Tradeoffs between throughput, buffer constraints, coding blocklength, and sensing duration for both fixed-rate and variable-rate transmissions are analyzed numerically. The relations between average error probability, sensing threshold and sensing duration are studied in the case of variable-rate transmissions.
\end{abstract}


\section{Introduction}

The main goal of the cognitive radio technology is to improve the efficiency in the use of limited, temporally and spatially under-utilized licensed radio frequency spectrum. A cognitive radio was first introduced by Mitola in \cite{mitola} as a smart wireless device, which senses the environment, learns and automatically adapts its transmission parameters without changing any hardware structure. Through such cognition and the reconfigurability features, cognitive radio systems enable cognitive users (unlicensed or secondary users) to perform spectrum sensing and access the channels based on the sensing results. Hence, the spectrum can be utilized opportunistically by allowing the cognitive users to either use the channel if there is no activity of primary users (licensed users) \cite{haykin} or share the spectrum with primary users under certain interference constraints. Motivated by the concept of a cognitive radio for efficient spectrum management, IEEE recently published IEEE 802.22 standard for wireless regional area networks (WRAN), which is the first cognitive radio based standard for using spectrum holes in TV broadcast bands by cognitive users \cite{IEEEstd}. It is required that cognitive users' transmission does not degrade the performance of primary users, such as TV users, through harmful interference.

The performance of cognitive radio systems has been extensively studied in order to obtain more insights regarding their potential applications. In particular, the performance limits of spectrum-sharing schemes were studied in \cite{ghasemi} by deriving the capacity of non-fading AWGN and fading channels under peak and average received-power constraints at the primary receiver. In addition to interference power constraints, peak and average transmit power constraints were taken into consideration in \cite{kang}, where the authors determined the optimal power allocation strategies for the ergodic and outage capacity of a secondary user fading channel under spectrum sharing system subject to different combinations of these constraints. In practical scenarios, errors in channel sensing are inevitable because of uncertainties in a communication channel, e.g., noise and fading. Therefore, the authors in \cite{kang2} considered the impact of imperfect channel sensing results on the ergodic capacity subject to average interference and transmit power constraints. Moreover, the outage capacity and truncated channel inversion with fixed rate (TIFR) capacity were studied in the presence of sensing errors in \cite{stotas}. The work in \cite{liang} investigated the optimal sensing duration that maximizes the achievable throughput of the secondary users. On the other hand, to overcome the problem of sensing-throughput tradeoff, the authors in \cite{stotas2}  proposed a novel cognitive radio system in which spectrum sensing and data transmission are performed at the same time by using the novel receiver structure based on perfect cancellation of the secondary signal. Recently, the authors in \cite{Xu} proposed optimal power allocation schemes to minimize the average bit error rate subject to peak/average transmit power and peak/average interference power constraints in spectrum sharing systems.

All of the above works assume the availability of perfect channel side information (CSI) of the interference channel between the secondary transmitter and primary receiver as well as the transmission channel between the secondary transmitter and the secondary receiver. However, in practice it is not an easy task to obtain perfect knowledge of the fading realizations. Therefore, the authors in \cite{kim}--\cite{smith} consider the capacity of cognitive radio systems under imperfect channel side information. In \cite{kim}, ergodic capacity was analyzed under average-received power and peak-received power constraints in the presence of only channel estimation error of the link between the secondary transmitter and primary receiver.  Another capacity metric, namely secondary user mean capacity, was investigated in \cite{suraweera} with partial CSI knowledge of the interference channel under a peak interference constraint. Recently, the authors in \cite{rezki} provided unified expressions of the ergodic capacity for different CSI level of the transmission link between the secondary transmitter and secondary receiver, and the interference link between the secondary transmitter and primary receiver subject to an average or a peak transmit power constraint together with an interference outage constraint. Different from these works, the authors in \cite{smith} also considered a minimum signal-to-interference noise ratio (SINR) constraint for the primary user and the interference from the primary transmitter on the secondary user mean capacity under different level of channel knowledge of the link between the primary transmitter and the primary receiver, and the link between the secondary transmitter and the primary receiver.

Another important consideration for cognitive radio systems especially in streaming and interactive multimedia applications is to support quality-of-service (QoS) requirements of secondary users in terms of buffer or delay constraints. In this respect, the authors in \cite{tang} obtained the optimal power adaptation policy to maximize the effective capacity subject to a given QoS constraint in multichannel communications.  In \cite{musavian2}, the optimal rate and power allocation strategy for the ergodic capacity in Nakagami fading channels was investigated under statistical delay QoS constraints. Moreover, the recent work in \cite{musavian3} mainly focused on the impact of adaptive $M$-QAM modulation on the effective capacity of secondary users under interference power and delay-QoS constraints.

Notably, in most studies as also seen in the aforementioned works, it is implicitly assumed that channel codes with arbitrarily long codewords can be used for transmission and consequently the well-known logarithmic channel capacity expressions of Gaussian channels are employed for analysis. In this paper, we depart from this idealistic assumption and assume that finite blocklength codes are used by the cognitive secondary users for sending messages. Hence, in our setup, transmission rates are possibly less than the channel capacity and errors can occur leading to retransmission requests. We further assume that the cognitive users operate under QoS constraints imposed as limitations on the buffer violation probability. The secondary users first detect the primary user activity, which is modeled as two-state Markov chain with busy and idle states. Subsequently, depending on the sensing result, the secondary user adapts its transmission power and rate and sends the data. Channel between the secondary users is assumed to be a block-fading channel in which the fading coefficient remains constant within each frame during the transmission. We first consider the scenario with perfect CSI at the secondary receiver and no CSI at the secondary transmitter. In this case, transmission is performed at two constant rate levels, depending on the sensing decision. In the second scenario, CSI is assumed to be available at both the secondary transmitter and receiver, enabling the secondary user to adapt its transmission rate according to the channel conditions. Under these assumptions, we analyze the throughput in the presence of buffer constraints by making use of the effective capacity formulation \cite{dapeng}, \cite{chang}, \cite{akin} and the recent results in \cite{yury}.

The analysis described above is conducted for a cognitive radio system model in which we have a single secondary transmitter, a single secondary receiver, and one or more primary users.

The rest of this paper is organized as follows: We introduce the system model in the next section. Section \ref{sec:preliminaries} provides preliminaries regarding the channel capacity with finite blocklength codes and effective throughput under statistical QoS constraints. In Section \ref{sec:effective_throughput}, we study effective throughput under the following two assumptions: CSI is known perfectly by either the receiver only or both the transmitter and receiver. The numerical results are presented and discussed in  Section \ref{sec:sim_results}, followed by conclusions in Section \ref{sec:conclusion}.

\section{System Model}
\label{sec:system_model}
In the cognitive radio model we consider, secondary users first determine the channel status (i.e., idle or busy) through spectrum sensing and then enter into data transmission phase with rate and power that depend on the sensing decision. Secondary users are allowed to coexist with primary users in the channel as long as their interference level does not deteriorate the performance of primary users. We also assume that channel sensing and data transmission are performed in frames of $T$ seconds. Duration of first $N$ seconds is allocated to channel sensing in which the secondary users observe either primary users' faded sum signal plus Gaussian background noise or just Gaussian background noise, and make a decision on primary user activity. In the remaining $T-N$ seconds, data transmission is performed over a flat-fading channel with additive Gaussian background noise and possibly additive interference arising due to  transmissions from active primary users.

\subsection{Markov Model for Primary User Activity}
\begin{figure}[htb]
\centering
\includegraphics[width=0.35\textwidth]{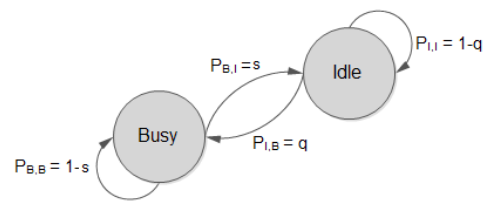}
\caption{Two-state Markov chain to model the primary user activity.}
\label{fig:transtion_primary}
\end{figure}

It is assumed that the primary users' activity in the channel remains the same during the frame duration of $T$ seconds. On the other hand, activity from one frame to another or equivalently the channel being busy or idle is modeled as a two-state Markov chain depicted in Figure \ref{fig:transtion_primary}. The busy state indicates that primary users are active in the channel whereas idle state represents no primary user activity. In Fig. \ref{fig:transtion_primary}, $P_{i, j}$, with $i, j \in \{I, B\}$, denotes the transition probability from state $i$ to state $j$, satisfying $\sum\limits_{j}P_{i,j} = 1$. Note that we set $P_{B, I} = s$ and $P_{I, B} = q$.

Given the above two-state Markov chain, we can easily determine the prior probabilities that the channel is busy and idle, denoted by $\Pr(\mH_{1})$ and $\Pr(\mH_{0})$, respectively, as follows:
\begin{equation}
\small
\begin{split}
\label{eq:prior_probs}
\Pr(\mH_{1}) = \frac{P_{I, B}}{P_{I, B} + P_{B, B}} = \frac{q}{q + s}, \hspace{0.1cm} \Pr(\mH_{0}) = \frac{P_{B, I}}{P_{B, I} + P_{I, I}} = \frac{s}{s + q}
\end{split}
\normalsize
\end{equation}
with notations $\mH_{0}$ and $\mH_{1}$ described below.
\subsection{Channel Sensing}
Channel sensing is performed in the first $N$ seconds. The remaining duration of $T - N$ seconds is reserved for data transmission. As in \cite{akin}, we formulate channel sensing as a binary hypothesis testing problem:
\begin{equation}
\begin{array}{ll}
\mH_{0} : y_i = n_i  &i=1,2,\dots , NB \\
\mH_{1} : y_i = s_i + n_i &i=1,2,\dots , NB
\end{array}
\label{eq:hypothesis}
\end{equation}
where $n_i$ denotes complex circularly symmetric background Gaussian noise samples with mean zero and variance $\E\{|n_i|^2\} = \sigma_n^2$, i.e., $n \sim \CN(0, \sigma_{n}^2)$. $s_i$ denotes the primary users' faded sum signal at the cognitive secondary receiver and can, for instance, be expressed as
\begin{gather}
s_i = \sum_{j = 1}^K g_{ps,j} \, u_j \label{eq:fadedsumsignal}
\end{gather}
where $K$ is the number of active primary transmitters, $u_j$ is the $j^{\text{th}}$ primary user's transmitted signal and $g_{ps,j}$ denotes the fading coefficient between the $j^{\text{th}}$ primary transmitter and the secondary receiver.
Therefore, hypothesis $\mH_{0}$ above corresponds to the case in which primary users are inactive in the channel whereas hypothesis $\mH_{1}$ models the presence of active primary users. Above, $B$ denotes the bandwidth of the system and therefore we have $NB$ complex signal samples in the sensing duration of $N$ seconds.

We further assume that $\{s_i\}$ is an independent and identically distributed (i.i.d.) sequence of circularly symmetric, complex Gaussian random variables with mean zero and variance $\E\{|s_i|^2\}= \sigma_s^2$, i.e., $s \sim \CN(0, \sigma_{s}^2)$. The optimal Neyman-Pearson energy detector is employed for channel sensing, and under the above-mentioned statistical assumptions, the test statistic is the total energy gathered in $N$ seconds, which is compared with a threshold $\lambda$:
\begin{equation}
\label{eq:NP_detector}
T(y) = \frac{1}{NB}\sum_{i=1}^{NB}|y_i|^2 \gtrless_{\mH_{0}}^{\mH_{1}} \lambda.
\end{equation}
Above, $T(y)$ is the sum of $NB$ independent $\chi^2$-distributed complex random variables and hence is itself $\chi^2$-distributed with $2NB$ degrees of freedom. With this characterization, the false alarm and detection probabilities can be expressed as
\begin{equation}
\small
\begin{split}
\label{false_alarm_probability}
P_f &= \Pr\{T(y) > \lambda | \mH_{0}\} = \Pr(\hH_{1}|\mH_{0}) = 1 - P\left( \frac{NB\lambda}{\sigma_n^2},  NB\right),
\end{split}
\normalsize
\end{equation}
\begin{equation}
\small
\begin{split}
\label{detection_probability}
P_d &= \Pr\{T(y) > \lambda | \mH_{1}\} = \Pr(\hH_{1}|\mH_{1}) = 1 - P\left( \frac{NB\lambda}{\sigma_n^2 + \sigma_s^2},  NB\right)
\end{split}
\normalsize
\end{equation}
where $P(s, x) = \frac{\gamma(s, x)}{\Gamma(s)}$ is the regularized Gamma function \cite[eq. 6.5.1]{abramowitz}, $\gamma(s, x)$ is the lower incomplete Gamma function \cite[eq. 6.5.2]{abramowitz}, and $\Gamma(s)$ is the Gamma function \cite[eq. 6.1.1]{abramowitz}. Additionally, $\hH_{1}$ and $\hH_{0}$ denote busy and idle sensing decisions, respectively. We further express the rest of the conditional probabilities of channel sensing decisions given channel true states, i.e., $\Pr(\hH_{i}|\mH_{j})$, in terms of $P_d$ and $P_f$, as follows:
\begin{align}
\label{eq:sensing_condprobs}
\Pr(\hH_{i}|\mH_{j}) = \begin{cases}
1 - P_d &\text{if } j = 1, i = 0\\
1 - P_f &\text{if } j = 0, i = 0
\end{cases}.
\end{align}
Combining (\ref{false_alarm_probability}) -- (\ref{eq:sensing_condprobs}) and applying the Bayes' rule, we can obtain the probabilities of channel being sensed to be busy and idle as
\begin{align}
\label{eq:busy_sensed_prob}
\Pr(\hH_{1}) &= \frac{q}{q + s}P_d + \frac{s}{s + q}P_f, \\
\label{eq:idle_sensed_prob}
\Pr(\hH_{0}) &= \frac{q}{q + s}(1 - P_d) + \frac{s}{s + q}(1 - P_f).
\end{align}

Finally, we would like to note that channel sensing can be performed by either the secondary receiver or transmitter, and we have implicitly assumed that the secondary receiver performs this task. In such a case, we further assume that the binary sensing decision made by the secondary receiver is reliably fed back to the secondary transmitter through a low-rate control channel.

\subsection{Data Transmission Parameters, Interference Management, and Channel Model} \label{subsec:channelmodel}

\subsubsection{Data Transmission Power and Rate}
Following channel sensing, secondary users initiate the data transmission phase in the remaining $T - N$ seconds. They adapt transmission rates and power levels depending on the channel sensing decision and availability of channel side information (CSI). More specifically, in the absence of CSI at the secondary transmitter, fixed-rate transmission is performed with constant power level while in the presence of perfect CSI, data is sent at a variable rate. Additionally, the average power is $\overline{P}_1$ and transmission rate is $r_1$ in the case of channel being sensed to be busy, and average power is $\overline{P}_2$ and transmission rate is $r_2$ in the case of channel being sensed to be idle.

\subsubsection{Interference Management}
The two-level transmission scheme described above is adopted to limit the interference inflicted on the primary users. Therefore, we in general have $\Pb_1 \le \Pb_2$. If cognitive users are not allowed to transmit when the primary user activity is detected in the channel, then we set $\Pb_1 = 0$. In general, power $\Pb_1$ should be below a certain threshold in order to limit the interference inflicted on the primary users. Note that when the transmission power is $\Pb_1$, the average interference experienced by a primary user is $\Pb_1 \E\{|g_{sp}|^2\}$ where $g_{sp}$ is the fading coefficient of the channel between the secondary transmitter and primary receiver. Then, an upper bound on the transmission power $\Pb_1$ can be expressed as
\begin{align}
\label{eq:interference_power_constraint}
\Pb_1 \le \frac{I_0}{\max_{j} \mathbb{E}\{|g_{sp,j}|^2\}}
\end{align}
where $I_0$ is the maximum average interference power that the primary users can tolerate and $|g_{sp,j}|^2$ is the channel gain between the secondary transmitter the and $j^{\text{th}}$ primary receiver. However, this may not provide sufficient protection in the presence of sensing errors since primary receivers are disturbed with average transmission power of $\Pb_2$ in the case of miss-detections. Therefore, as an additional mechanism to control the interference, an upper bound on the probability of miss detection or equivalently a lower bound on the detection probability should be imposed in cognitive radio systems so that miss-detections occur rarely.

Yet, another method to limit the average interference power experienced by the primary users is to impose the following constraint on the transmission powers:
\begin{align} \label{eq:avg-interference-power}
P_d \Pb_1 + (1-P_d) \Pb_2 \le \frac{I_0}{\max_{j} \mathbb{E}\{|g_{sp,j}|^2\}}
\end{align}
together with possibly peak constraints $\Pb_1 \le \Pb_{\text{peak,1}}$ and $\Pb_2 \le \Pb_{\text{peak,2}}$.  Above, $P_d$ is the detection probability in channel sensing. Note that primary receiver is disturbed with transmissions of power $\Pb_1$ and $\Pb_2$ with probabilities $P_d$ and $(1-P_d)$, respectively, which are the probabilities of correct detection and miss-detection events. Hence, average interference power is proportional to $P_d \Pb_1 + (1-P_d) \Pb_2$. We note that such an average interference power constraint was, for instance, considered in \cite{kang2}.

Finally, we remark that the analysis in Section \ref{sec:effective_throughput} is conducted for given average power constraints and given signal-to-noise ratios. Therefore, any of the interference constraints discussed above can be easily be accommodated in the subsequent throughput analysis.

\subsubsection{Channel Model}
Next, we describe the channel model. The channel between the secondary users is assumed to experience flat fading. We also consider the block-fading assumption in which the fading coefficients are constant within the frame of $T$ seconds and change independently between the frames.  Under these assumptions, the complex input-complex output relationship is
\begin{align}
\label{eq:io_relations}
\y = \begin{cases}
h\x + \n &\text{in the absence of primary user activity,}\\
h\x + \n + \s &\text{in the presence of primary user activity}.
\end{cases}
\end{align}

Above, $h$ is the circularly-symmetric complex fading coefficient with a finite variance, i.e., $\E\{|h|^2\} < \infty$. $\x$ and $\y$ are the $(T-N)B$--dimensional complex channel input and output vectors, respectively. Since we assume that transmissions are power constrained by $\Pb_1$ or $\Pb_2$, the average energy available in the data transmission period of $(T-N)$ seconds is $(T-N)\Pb_i$ for $i = 1,2$, and hence
$\E\{\|\x\|^2\} = (T-N)\Pb_i$. With energy uniformly distributed across input symbols, the average energy per symbol becomes $\E\{|x_i|^2\} = \frac{\Pb_i}{B}$\footnote{Alternatively, if an average energy constraint of $\E\{\|\x\|^2\} = \Pb_i T$ is imposed in the data transmission period rather than an average power constraint, the average energy per symbol becomes $\E\{|x_i|^2\} = \frac{T\Pb_i}{(T-N)B}$. This leads to the scaling of the signal-to-noise ratio by a factor of $\frac{T}{T-N}$. Since the analysis in Section \ref{sec:effective_throughput} is conducted for given signal-to-noise ratio expressions, an average energy constraint given as above can be incorporated into the analysis easily.}.

In (\ref{eq:io_relations}), $\n$ denotes the vector of i.i.d. noise samples that are circularly symmetric, Gaussian random variables with mean zero and variance $E\{|n|^2\}=\sigma_n^2$, and $\s$ again represents the vector of active primary users' faded sum signal received at the secondary receiver similarly as in (\ref{eq:fadedsumsignal}). We again assume that the components of $\s$ are i.i.d. Gaussian random variables with mean zero and variance $E\{|s|^2\}=\sigma_s^2$.

\section{Preliminaries}
\label{sec:preliminaries}
In this section, we briefly review rates achieved with finite blocklength codes and effective throughput under statistical QoS constraints.
\subsection{Transmission Rate in the Finite Blocklength Regime} \label{subsec:finiteblocklength}

In \cite{yury}, Polyanskiy, Poor and Verd\'u studied the channel coding rate achieved with finite blocklength codes and identified a second-order expression for the channel capacity of the real additive white Gaussian noise (AWGN) channel in terms of the coding blocklength $(T - N)B$, error probability $\e$, and signal-to-noise ratio (SNR). As done in \cite{gursoy-icc11}, this result can be slightly modified to obtain the following approximate expression for the instantaneous channel capacity of a flat-fading channel attained in the data transmission duration of $(T-N)B$ symbols\footnote{For (\ref{eq:codingratef}) to hold, we assume that $(T-N)B$ is sufficiently large but finite.}:
\begin{equation}
\small
\begin{split}
\label{eq:codingratef}
r = \log_2(1+&\ssnr|h|^2) \\&- \sqrt{\frac{1}{(T-N)B} \left( 1- \frac{1}{(\ssnr|h|^2 + 1)^2} \right)} Q^{-1}(\e) \log_2 e
\end{split}
\normalsize
\end{equation}
where $Q(x) = \int_x^{\infty}\frac{1}{\sqrt{2\pi}}e^{-t^2/2}dt$ is the Gaussian $Q$-function and $\ssnr$ denotes the signal-to-noise ratio which can be expressed as average energy per symbol normalized by the variance of the noise random variable. The above expression provides the rate that can be achieved with error probability $\e$ for a given fading coefficient $h$ and signal-to-noise ratio SNR. Note that as the blocklength $(T - N)B$ grows without bound, the second term on the right-hand side of (\ref{eq:codingratef}) vanishes and transmission rate $r$ approaches the instantaneous channel capacity $\log_2(1+\ssnr|h|^2)$ for any arbitrarily small $\e > 0$.

Equivalently, we can also conclude from (\ref{eq:codingratef}) that transmission with a given fixed rate $r$ can be supported with error probability
\begin{gather}
\label{eq:errorprob}
\e_{|h|^2} = Q\left( \frac{\log_2(1 + \ssnr|h|^2) - r}{\sqrt{\frac{1}{(T - N)B}\left( 1 - \frac{1}{(\ssnr|h|^2 + 1)^2}\right)}\log_2e} \right)
\end{gather}
where the dependence of the error probability on fading is made explicit by expressing $\e$ with subscript $|h|^2$.

\begin{figure}[htb]
\centering
\includegraphics[width=\figsize\textwidth]{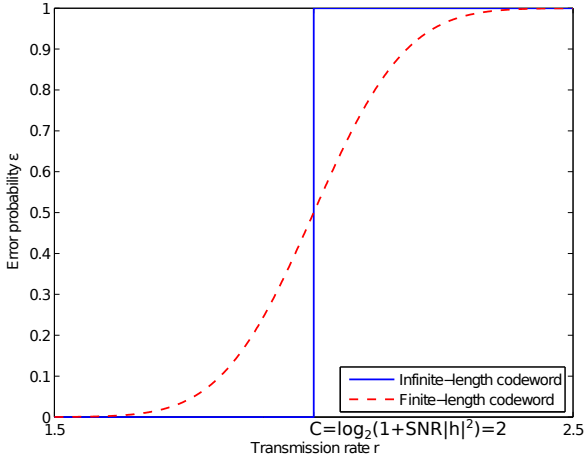}
\caption{Error probability vs. transmission rate for infinite-length and finite-length codewords, $\ssnr=3, |h|^2=1$, and $C = \log_2(1+\ssnr|h|^2) = 2$.}
\label{fig:E_vs_r}
\end{figure}

In order to observe the effect of finite-length codewords on the reliability of transmissions, in Fig. \ref{fig:E_vs_r} we display the error probability vs. transmission rate when the transmitter is assumed to employ finite-length codewords together with the asymptotical behavior as the codeword length grows without bound. According to the Shannon capacity limit, when the codeword length increases without bound, we can achieve reliable transmission with no decoding errors (i.e.,  $\epsilon=0$) for any transmission rate less than the instantaneous channel capacity, i.e., $r < C = \log_2(1+\ssnr|h|^2)$, whereas  reliable communication is not possible when $r \ge C$. Indeed, as noted in \cite{info-theory-book}, by the strong converse, when $r >C$, probability of error goes exponentially to 1 as the blocklength increases. Therefore, we have the sharp cutoff at the instantaneous capacity in Fig. \ref{fig:E_vs_r} for the asymptotic scenario of codewords of infinite length. Close inspection of (\ref{eq:errorprob}) leads to the same conclusion as well. Let $r > \log_2(1 + \ssnr|h|^2)$. Then, as the blocklength $(T - N)B$ increases to infinity, the term $\sqrt{\frac{1}{(T - N)B}\left( 1 - \frac{1}{(\ssnr|h|^2 + 1)^2}\right)}$ vanishes and in the limit, we have $\e_{|h|^2} = Q(-\infty) = 1$. If $r < \log_2(1 + \ssnr|h|^2)$, we asymptotically have $\e_{|h|^2} = Q(\infty) = 0$.

On the other hand, for finite-length codewords, when we plot (\ref{eq:errorprob}), we see that we have a relatively smooth transition. This behavior indicates that for transmissions with rates less than the instantaneous capacity, we can still have errors, albeit with relatively small probabilities, while transmission rates above the instantaneous capacity can lead to successful transmissions but again only with small probability.

\subsection{Throughput under Buffer Limitations}

In \cite{dapeng}, Wu and
Negi  defined the effective capacity as the maximum
constant arrival rate that a given service process can support in
order to guarantee a statistical QoS requirement characterized by the
QoS exponent $\theta$. If we define $\Q$ as
the stationary queue length, then $\theta$ is the decay rate of the
tail of the distribution of the queue length $\Q$:
\begin{equation}
\label{eq:decay_rate}
\lim_{q \to \infty} \frac{\log P(\Q \ge q)}{q} = -\theta.
\end{equation}
Therefore, for large $q_{max}$, the buffer overflow probability can be approximated as exponentially decaying at a rate $\theta$:
\begin{gather}
\label{eq:bufferconstraint}
P(\Q \ge q_{\max}) \approx
e^{-\theta q_{\max}}.
\end{gather}
Hence, larger values of $\theta$ represent more strict QoS constraints whereas lower values of $\theta$ indicate looser QoS guarantees.

The effective capacity, which quantifies the throughput under a buffer constraint in the form of (\ref{eq:bufferconstraint}), is given by (\cite{dapeng}, \cite{chang})
\begin{gather}
\label{eq:eff-cap-def}
R_E = -\lim_{t\rightarrow\infty}\frac{1}{\theta
t}\log_e{\mathbb{E}\{e^{-\theta S[t]}\}} \triangleq -\frac{\Lambda(-\theta)}{\theta}
\end{gather}
where $\Lambda(\theta) = \lim_{t\rightarrow
\infty}\frac{1}{t}\log_{e}\mathbb{E}\{e^{\theta S[t]}\}$,  $S[t] = \sum_{i=1}^{t}r_i$ is the time-accumulated service
process and $\{r_i, i=1,2,\ldots\}$ denotes the discrete-time
stationary and ergodic stochastic service process. In the remainder of the paper, $R_E$ will be referred as the effective rate rather than the effective capacity since we study the performance in the finite blocklength regime.

\subsection{Impact of Finite-Blocklength Analysis in Cognitive Radio Channels under Buffer Limitations}

Before a detailed analysis, we in this subsection briefly describe the impact of considering finite-blocklength regime in the throughput analysis of cognitive radio channels in the presence of buffer constraints. As pointed out in Section \ref{subsec:finiteblocklength}, the critical difference from the studies with infinite-blocklength codes is that we now have non-zero error probabilities even if the transmission rates are less than the instantaneous capacity. Moreover, we observe from (\ref{eq:errorprob}) that error probabilities, for fixed-rate transmissions, fluctuate depending on the channel conditions.  In general, such error events will be reflected in the subsequent analysis by the presence of OFF states in which reliable communication is not achieved due to errors and consequently retransmissions are required. This potentially has significant impact in buffer-limited systems as frequent communication failures and retransmission requests can easily lead to buffer overflows. Therefore, coding rates and error probabilities in the finite-blocklength regime should be judiciously analyzed and optimal transmission parameters should be identified. Situation is further exacerbated in cognitive radio systems in which channel sensing is performed imperfectly and interference constraints are imposed. Firstly, time allocated to channel sensing results in reduced transmission duration, leading to reduced codeword blocklength with consequences on both the rates and error probabilities. Additionally,  false-alarms and miss-detections, experienced due to imperfect sensing, cause over- or underestimations of the channel, and resulting mismatches cause transmission rates and/or error probabilities to exceed or be lower than required or target levels (for instance, as will be discussed in Section \ref{sec:state_transition_perfectCSI}).

\section{State Transition Model for the Cognitive Radio Channel and Effective Throughput}
\label{sec:effective_throughput}
In this section, we first construct an eight-state Markov chain in order to model the cognitive radio channel, and then derive the corresponding state transition probabilities when CSI is assumed to be perfectly known either at the receiver only or at both the receiver and transmitter. Subsequently, we analyze the throughput achieved with finite blocklength codes in the presence of buffer constraints under these two assumptions.
\subsection{Perfect CSI at the Receiver Only}
\label{sec:fixed_rate}
It is assumed that perfect knowledge of fading realizations is available at the secondary receiver, but not at the secondary transmitter. Therefore, the transmitter performs data transmission with constant rate of   $r_1$ or $r_2$ based on the sensing decision about the channel occupancy by the primary users.

\subsubsection{State Transition Model}\label{sec:state-transition}
Before analyzing the throughput achieved by the secondary users with finite blocklength codes  under buffer constraints, we construct a state transition model for the cognitive radio channel. First, we list the four possible scenarios, together with corresponding signal-to-noise ratio expressions, arising as a result of different channel sensing decisions and the true nature of primary users' activity:
\begin{itemize}
\item  \emph{Scenario \Rmnum{1} } (Correct-detection denoted by joint event $(\mH_{1}, \hH_{1})$):\\Busy channel is sensed as busy and $\tSNR_1 = \frac{\overline{P}_1}{B (\sigma_n^2 + \sigma_{s}^2)}$.
\item  \emph{Scenario \Rmnum{2} } (Miss-detection denoted by $(\mH_{1}, \hH_{0})$):\\Busy channel is sensed as idle and $\tSNR_2 = \frac{\overline{P}_2}{B (\sigma_n^2 + \sigma_{s}^2)}$.
\item  \emph{Scenario \Rmnum{3} } (False-alarm denoted by $(\mH_{0}, \hH_{1})$):\\ Idle channel is sensed as busy and $\tSNR_3 = \frac{\overline{P}_1}{B\sigma_n^2}$.
\item  \emph{Scenario \Rmnum{4} } (Correct-detection denoted by $(\mH_{0}, \hH_{0})$):\\ Idle channel is sensed as idle and $\tSNR_4 = \frac{\overline{P}_2}{B\sigma_n^2}$.
\end{itemize}

\begin{figure}[htb]
\centering
\includegraphics[width=0.45\textwidth]{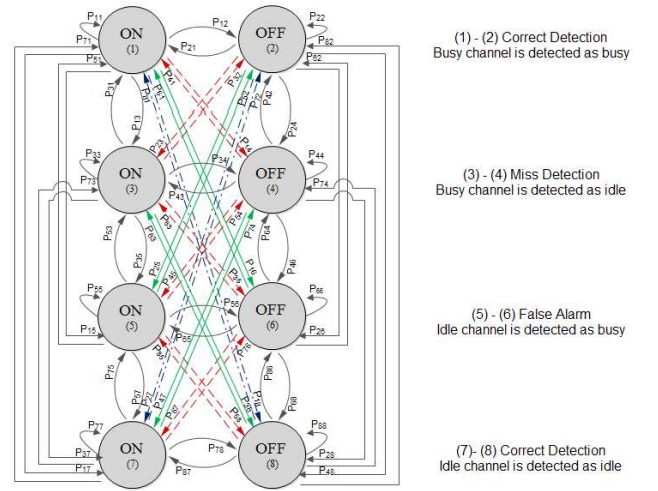}
\caption{The state-transition model for the cognitive radio channel with eight possible states.}
\label{fig:transtion_cognitive}
\end{figure}

Additionally, transmission rate is $r_1$ bits/s/Hz in scenarios 1 and 3 above, and is $r_2$ bits/s/Hz in scenarios 2 and 4. When codewords of length $(T-N)B$ are used to send the data at these fixed rates, we know from the discussion in Section \ref{subsec:finiteblocklength} that information is received reliably with probability $(1-\e_{|h|^2})$ while errors occur and retransmission is needed with probability $\e_{|h|^2}$ as formulated in (\ref{eq:errorprob}). More specifically, the error probabilities in scenarios 1 and 3 are
\begin{align}
\label{eq:error_13}
\e_d(|h|^2)= Q\left( \frac{\log_2(1 +\ssnr_d|h|^2) - r_1}{\sqrt{\frac{1}{(T -N)B}\left( 1 - \frac{1}{(\ssnr_d|h|^2 + 1)^2}\right)}\log_2e} \right)
\end{align}
for $d = 1$ and 3, respectively. Similarly, we have
\begin{align}
\label{eq:error_24}
\e_l(|h|^2) = Q\left( \frac{\log_2(1 + \ssnr_l|h|^2) - r_2}{\sqrt{\frac{1}{(T - N)B}\left( 1 - \frac{1}{(\ssnr_l|h|^2 + 1)^2}\right)}\log_2e} \right)
\end{align}
in scenarios 2 and 4 for $l = 2$ and 4, respectively. Above, we see that error probability is a function of the fading coefficient $|h|$ and SNR. From this discussion, we conclude that the channel can be either in the ON state (in which information is reliably received) or the OFF state (in which erroneous reception occurs) in each scenario. Hence, we have eight states in total in the Markov model for the cognitive radio channel as depicted in Fig. \ref{fig:transtion_cognitive}.
Note that since reliable communication cannot be achieved in the OFF states, the transmission rate is effectively zero and the data has to be retransmitted in these states.  Therefore, the service rates (in bits/frame) in four scenarios can be expressed, respectively, as
\begin{align}
\label{eq:servicerate1}
R_d = \begin{cases} 0 & \text{with probability } \e_d(|h|^2)  \\
 (T - N)Br_1 & \text{with probability } (1 - \e_d(|h|^2))
\end{cases} \\
R_l= \begin{cases} 0 & \text{with probability } \e_l(|h|^2)  \\
 (T - N)Br_2 & \text{with probability } (1 - \e_l(|h|^2))
\end{cases}
\end{align}
for  $d= 1, 3$ and $l = 2, 4$.

Next, we identify the transition probabilities from state $i$ to state $k$ denoted by $p_{ik}$ in the eight state transition model of the cognitive radio channel. We initially analyze in detail $p_{11}$, the probability of staying in the topmost ON state.
\begin{figure*}
\small
\begin{align}
\label{main_probability}
p_{11} = &\Pr\left\{\begin{array}{cc}\begin{array}{c}\text{Channel is busy and detected as busy}\\\text{and channel is ON} \text{ in the } i^{th} \text{ frame}\end{array} \Big\vert&\begin{array}{c} \text{Channel was busy and detected as busy}\\
\text{and channel was ON in the } (i - 1)^{th} \text{ frame}\end{array} \end{array} \right\}\\
\label{seperated_probability}
= &\Pr\left\{\begin{array}{cc} \begin{array}{c} \text{Channel is busy}\\ \text{in the } i^{th} \text{ frame}\end{array} \Big\vert&\begin{array}{c}\text{Channel was busy}\\ \text{in the } (i - 1)^{th} \text{ frame}\end{array} \end{array} \right\} \times \Pr\left\{\begin{array}{cc} \begin{array}{c} \text{Channel is detected as busy}\\ \text{in the } i^{th} \text{ frame}\end{array} \Big\vert&\begin{array}{c} \text{Channel is busy}\\ \text{in the } i^{th} \text{ frame}\end{array} \end{array} \right\}\nonumber\\
&\times \Pr(\text{Channel is ON in the } i^{th} \text{ frame})
\\
= &(1 - s)P_d(1- \e_1(|h|^2) \label{separated_probability4}
\end{align}
\normalsize
\hrule
\end{figure*}
We can first express $p_{11}$ as in (\ref{main_probability}) shown at the top of the next page. Subsequently, we can write (\ref{seperated_probability}) by noting that channel being actually busy in the current frame depends on its state in the previous frame due to the two-state Markov chain, and
channel being detected as busy in the $i^{th}$ frame depends only on the true state of the channel being busy or idle in the $i^{th}$ frame and not on previous true states and sensing decisions since channel sensing is performed in each frame independently. Moreover, channel being ON does not depend on the sensing decisions and channel being ON or OFF in the previous frames due to the block-fading assumption. Finally, we have (\ref{separated_probability4}) by observing that the first probability in (\ref{seperated_probability}) is $P_{B, B} = 1-s$ in the Markov chain, the second probability is the correct detection probability $P_d$ in channel sensing, and channel is ON with probability $(1-\epsilon_1(|h|^2))$ as discussed above.\\
By following the same steps, transition probabilities from all eight states to state 1 can be found as
\begin{equation}
\begin{split}
\label{eq:prob1} p_{i1} &= p_{11} = p_{21} = p_{31} = p_{41} = (1 - s)P_d(1-\e_1(|h|^2)), \\
p_{k1} &= p_{51} = p_{61} = p_{71} = p_{81} = qP_d(1-\e_1(|h|^2)).
\end{split}
\end{equation}
The channel is busy in the first four states and we see that the transition probabilities from these four states to the first state are the same. The channel is idle in the last four states and similarly their transition probabilities are equal. Hence, (\ref{eq:prob1}) shows that we can group the transition probabilities into two with respect to the true nature of the channel, i.e., busy or idle.  The rest of the transition probabilities between each state can be derived in a similar fashion and the overall result can be listed as follows for $i = 1,2,3,4$ and $k = 5,6,7,8$:
\begin{equation}
\small
\begin{split}
\label{eq:probs}
\hspace{-0.2cm}\begin{array}{ll}
p_{i2} \!=\! (1 \!-\! s)P_d\e_1(|h|^2) & p_{k2} \!=\! qP_d\e_1(|h|^2),\\
p_{i3} \!=\! (1 \!-\! s)(1 \!-\! P_d)(1\!-\!\e_2(|h|^2))& p_{k3} \!=\! q(1 \!-\! P_d)(1\!-\!\e_2(|h|^2)),\\
p_{i4} \!=\! (1 \!-\! s)(1 \!-\! P_d)\e_2(|h|^2) & p_{k4} \!=\! q(1 \!-\! P_d)\e_2(|h|^2),\\
p_{i5} \!=\! sP_f(1\!-\!\e_3(|h|^2)) & p_{k5} \!=\! (1 \!-\! q)P_f(1\!-\!\e_3(|h|^2)),\\
p_{i6} \!=\! sP_f\e_3(|h|^2) & p_{k6} \!=\! (1 \!-\! q)P_f\e_3(|h|^2),\\
p_{i7} \!=\! s(1 \!-\! P_f)(1\!-\!\e_4(|h|^2)) & p_{k7} \!=\! (1 \!-\! q)(1 \!-\! P_f)(1\!-\!\e_4(|h|^2)),\\
p_{i8} \!=\! s(1 \!-\! P_f)\e_4(|h|^2) & p_{k8} \!=\! (1 \!-\! q)(1 \!-\! P_f)\e_4(|h|^2).
\end{array}
\end{split}
\normalsize
\end{equation}
The set of transition probabilities can be expressed in an $8 \times 8$ state transition matrix
\begin{align}
\label{state_matrix}
\hspace{-0.4cm}R=\left[\begin{array}{cccc}p_{1,1}&p_{1,2}&\dots&p_{1,8}\\ .&.&\dots&.\\p_{4,1}&p_{4,2}&\dots&p_{4,8}\\p_{5,1}&p_{5,2}&\dots&p_{5,8}\\.&.&\dots&.\\p_{8,1}&p_{8,2}&\dots&p_{8,8}\end{array}\!\right]\!=\!\left[\begin{array}{cccc}p_{i1}&.&.&p_{i8}\\ .&.&.&.\\p_{i1}&.&.&p_{i8}\\p_{k1}&.&.&p_{k8}\\.&.&.&.\\p_{k1}&.&.&p_{k8}\end{array}\!\right]. \end{align}
Note that the rank of $R$ is 2 since it has only two linearly independent column vectors.

\subsubsection{Throughput Under Buffer Limitations}
\label{sec:throughput_underQoS_noCSI}
In this subsection, we determine the throughput achieved with finite blocklength codes subject to buffer constraints by obtaining the effective rate of the cognitive radio channel with the state-transition model constructed in Section \ref{sec:state-transition}. The approach and techniques in this section closely follow \cite{akin} with the difference that we now consider performance in the finite blocklength regime. In \cite[Chap. 7, Example 7.2.7]{chang-book}, it is
shown for Markov modulated processes that
\begin{gather}
\label{eq:theta-envelope}
\frac{\Lambda(\theta)}{\theta} = \frac{1}{\theta} \log_e
sp(\phi(\theta)R)
\end{gather}
where $\Lambda(\theta)$ is defined underneath (\ref{eq:eff-cap-def}), $sp(\phi(\theta)R)$ is the spectral radius or the maximum of the absolute values of the eigenvalues of the matrix
$\phi(\theta)R$, $R$ is the transition matrix of the underlying Markov process, and $\phi(\theta) = \text{diag}(\phi_1(\theta), \ldots, \phi_M(\theta))$ is a diagonal matrix whose components are the moment generating functions of the processes in $M$ states ($M = 8$ in our model).  In our case, we have
\begin{align}
\label{eq:diagonal_matrix}
\phi(\theta) = \text{diag}\{&e^{\theta(T - N)Br_1},1,e^{\theta(T - N)Br_2},
1, \nonumber
\\
&e^{\theta(T - N)Br_1},1,e^{\theta(T - N)Br_2},1\}, \text{ and }
\end{align}

\begin{align}
\label{eq:phi_theta_matrix}
\phi(\theta)R=\left[\begin{array}{ccc}e^{\theta(T - N)Br_1}p_{i1}&\dots&e^{\theta(T - N)Br_1}p_{i8}\\
p_{i1}&\dots&p_{i8}\\
e^{\theta(T - N)Br_2}p_{i1}&\dots&e^{\theta(T - N)Br_2}p_{i8}\\
p_{i1}&\dots&p_{i8}\\
e^{\theta(T - N)Br_1}p_{k1}&\dots&e^{\theta(T - N)Br_1}p_{k8}\\p_{k1}&\dots&p_{k8}\\e^{\theta(T - N)Br_2}p_{k1}&\dots&e^{\theta(T - N)Br_2}p_{k8} \\p_{k1}&\dots&p_{k8}\end{array}\right].
\end{align}

Note that $\phi(\theta)R$ is a rank-2 matrix as well. As the $n$-rowed ($n\geq3$) principal minors of $\phi(\theta)R$ are zero, the coefficients of the characteristic polynomial of the matrix $\phi(\theta)R$ can be found in terms of adding only the 1-rowed and 2-rowed principal minors, then the maximum root of this polynomial gives the spectral radius $sp(\phi(\theta)R)$, which is expressed in (\ref{eq:sp}) on the next page. Now, combining (\ref{eq:eff-cap-def}), (\ref{eq:theta-envelope}), and (\ref{eq:sp}), we can easily express the effective rate of the cognitive channel as in (\ref{eq:eff-cap-cognitive}) on the next page. Note that $R_E(\tSNR, \theta)$ in (\ref{eq:eff-cap-cognitive}) characterizes the maximum constant arrival rates that the cognitive radio channel can support in the finite blocklength regime under buffer limitations characterized by the QoS exponent $\theta$. Note that this throughput is maximized over transmission rates $r_1$ and $r_2$.

Throughput in the absence of any buffer constraints, which can be easily determined by letting $\theta \to 0$ in $R_E(\tSNR, \theta)$, is given in (\ref{eq:eff-cap-theta-zero}).

\begin{figure*}
\begin{equation}
\label{eq:sp}
\begin{split}
sp(\phi(\theta)R) = &\frac{1}{2}\Big[\phi_1(\theta)p_{i1} + \dots + \phi_4(\theta)p_{i4} + \phi_5(\theta)p_{k5} + \dots + \phi_8(\theta)p_{k8} \Big]
\\&+ \frac{1}{2}\Bigg\{\Big[\phi_1(\theta)p_{i1} + \dots + \phi_4(\theta)p_{i4} - \phi_5(\theta)p_{k5} - \dots - \phi_8(\theta)p_{k8}\Big]^2\\&+ 4(\phi_1(\theta)p_{k1} + \dots + \phi_4(\theta)p_{k4})\times(\phi_5(\theta)p_{i5} + \dots + \phi_8(\theta)p_{i8})\Bigg\}^{\frac{1}{2}}
\end{split}
\end{equation}
\hrule
\end{figure*}

\begin{figure*}
\begin{equation}
\label{eq:eff-cap-cognitive}
\begin{split}
R_E(\tSNR, \theta) = &\max_{r_1, r_2\geq0}-\frac{1}{\theta TB}\log_e\mathbb{E}_{|h|^2}\Bigg(\frac{1}{2}\Big[(p_{i1} + p_{k5})e^{-\theta (T - N)Br_1} + (p_{i3} + p_{k7})e^{-\theta (T - N)Br_2} + p_{i2} + p_{i4}+ p_{k6} + p_{k8}\Big]
\\+ &\frac{1}{2}\Big\{\Big[(p_{i1} - p_{k5})e^{-\theta (T - N)Br_1} + (p_{i3} - p_{k7})e^{-\theta (T - N)Br_2} + p_{i2} + p_{i4}- p_{k6} - p_{k8}\Big]^2\\+ &4(p_{k1}e^{-\theta (T - N)Br_1} + p_{k2} + p_{k3}e^{-\theta (T - N)Br_2} + p_{k4})\times(p_{i5}e^{-\theta (T - N)Br_1} + p_{i6} + p_{i7}e^{-\theta (T - N)Br_2} + p_{i8})\Big\}^{\frac{1}{2}}\Bigg)
\end{split}
\end{equation}
\hrule
\end{figure*}

\begin{figure*}
\begin{equation}
\label{eq:eff-cap-theta-zero}
\begin{split}
R_E(\tSNR, 0) = \max_{r_1, r_2\geq0}&\frac{(T - N)r_1P_d}{2T(s + q)}\Big((1 - s)(3q - s) + 4sq\Big)(1 - \mathbb{E}_{\vert h \vert^2}\{\e_1(|h|^2)\}) \\&+ \frac{(T - N)r_2(1 - P_d)}{2T(s + q)}\Big((1 - s)(3q - s) + 4sq\Big)(1 - \mathbb{E}_{\vert h \vert^2}\{\e_2(|h|^2)\})\\&+\frac{(T - N)r_1P_f}{2T(s + q)}\Big((1 - s)(3s - q) + 4sq\Big)(1 - \mathbb{E}_{\vert h \vert^2}\{\e_3(|h|^2)\})\\&+ \frac{(T - N)r_2(1 - P_f)}{2T(s + q)}\Big((1 - s)(3s - q) + 4sq\Big)(1 - \mathbb{E}_{\vert h \vert^2}\{\e_4(|h|^2)\})
\end{split}
\end{equation}
\hrule
\end{figure*}

\subsection{Perfect CSI at both the Receiver and Transmitter}
Instead of CSI known by the receiver only, we in this section consider that both the secondary transmitter and receiver have access to perfect CSI. Therefore, in contrast to Section \ref{sec:fixed_rate}, the secondary transmitter can adapt its transmission scheme by varying the rate depending on the instantaneous values of the fading coefficient $|h|$.
\subsubsection{State Transition Model}
\label{sec:state_transition_perfectCSI}
Under the assumption of perfect CSI at the transmitter, the eight-state Markov model for the cognitive radio channel with four possible scenarios and ON/OFF states is unchanged as we defined in Section \ref{sec:state-transition}. Additionally, the SNR expressions in each scenario are still the same. In contrast to fixed-rate transmission schemes, for a given fixed target error probability $\epsilon$, the secondary transmitter now varies its transmission rate according to the channel conditions and channel sensing decision. More specifically, in the case of channel being sensed as busy, the secondary transmitter initiates data transmission with rate
\begin{equation}
\small
\begin{split}
\label{eq:r1_epsilon}
r_1(\ssnr_1, |h|^2) &= \log_2(1 +\ssnr_1|h|^2) \\&- \sqrt{\frac{1}{(T \!-\!N)B}\left(\!1 - \frac{1}{(\ssnr_1|h|^2 + 1)^2}\!\right) } Q^{-1}(\e)\log_2e.
\end{split}
\normalsize
\end{equation}
On the other hand, if no primary user activity is sensed in the channel, we have the following transmission rate
\begin{equation}
\small
\begin{split}
\label{eq:r2_epsilon}
r_2(\ssnr_4, |h|^2) &= \log_2(1 +\ssnr_4|h|^2) \\&- \sqrt{\frac{1}{(T \!-\!N)B}\!\left(1 - \frac{1}{(\ssnr_4|h|^2 + 1)^2}\!\right) } Q^{-1}(\e)\log_2e.
\end{split}
\normalsize
\end{equation}
Before specifying the transition probabilities of the cognitive radio channel, we initially determine the error probabilities in each scenario that are associated with the transmission rates $r_1(\ssnr_1, |h|^2)$ or $r_2(\ssnr_4, |h|^2)$:
\begin{itemize}
\item In scenario 1, the fixed target error probability $\e$ is attained with the transmission rate $r_1(\ssnr_1, |h|^2)$ defined above.
\item In scenario 2 (in which we have missed detection), due to the primary user activity and the resulting interference on secondary users, the actual channel rate associated with error probability $\e$ is
\begin{equation}
\small
\begin{split}
\label{eq:rate_scenario2}
\log_2(1 +\ssnr_2|h|^2) - \sqrt{\frac{1}{(T -N)B}\left(1 -\frac{1}{(\ssnr_2|h|^2 + 1)^2}\right) }\\ \times Q^{-1}(\e)\log_2e.
\end{split}
\normalsize
\end{equation}
However, the secondary users do not know the true state of the channel, and they only have the imperfect channel sensing result. In this case, the channel is detected as idle even if the primary users are active.  Hence, for the given target error probability $\e$, the secondary users send data with rate $r_2(\ssnr_4, |h|^2)$, which is obviously higher than the actual rate in ($\ref{eq:rate_scenario2}$) that the channel actually supports with error probability $\e$.

As a result, we have in fact higher error probability $\e''_{|h|^2}$ (compared to the given target error probability $\e$) when the transmission rate is $r_2(\ssnr_4, |h|^2$). Equating the transmission rate $r_2(\ssnr_4, |h|^2)$ to that in ($\ref{eq:rate_scenario2}$), and rearraging the terms, the final expression of the actual error probability $\e''_{|h|^2}$ can be found as (\ref{eq:r1_epsilondoublehead}) shown at the top of next page.
\begin{figure*}
\begin{align}
\label{eq:r1_epsilondoublehead}
\e''_{|h|^2} = Q \left( \frac{\log_2(\frac{1 +\ssnr_2|h|^2}{1 +\ssnr_4|h|^2}) + \sqrt{\frac{1}{(T -N)B}\left(1 - \frac{1}{(\ssnr_4|h|^2 + 1)^2}\right) } Q^{-1}(\e)\log_2e}{\sqrt{\frac{1}{(T -N)B}\left(1 - \frac{1}{(\ssnr_2|h|^2 + 1)^2}\right) }\log_2e}\right).
\end{align}
\hrule
\end{figure*}
In this case, due to the sensing error, we are subject to more transmission errors resulting in lower reliability in data transmission. We also see that error probability $\e''_{|h|^2}$ in (\ref{eq:r1_epsilondoublehead}) that can be achieved with transmission rate $r_2(\ssnr_4, |h|^2)$ is a function of the fading coefficient $|h|$.
\item In scenario 3 (in which we have false alarm), for a given error probability $\e$, the channel supports the rate
\begin{equation}
\small
\begin{split}
\label{eq:rate_scenario3}
\log_2(1 +\ssnr_3|h|^2) - \sqrt{\frac{1}{(T -N)B}\left(1 - \frac{1}{(\ssnr_3|h|^2 + 1)^2}\right) } \\ \times Q^{-1}(\e)\log_2e
\end{split}
\normalsize
\end{equation}
which is higher than the rate $r_1(\ssnr_1, |h|^2)$ because there is actually no interference from the primary users, i.e, $\tSNR_1 <  \tSNR_3$. Therefore, the error probability that can be attained with this transmission rate is less than the given fixed target error probability $\e$. Following the same approach adopted in scenario 2, the actual error probability $\e'_{|h|^2}$ can be expressed as (\ref{eq:r1_epsilonhead}) shown at the top of next page.
\begin{figure*}
\begin{align}
\label{eq:r1_epsilonhead}
\e'_{|h|^2} = Q \left( \frac{\log_2(\frac{1 +\ssnr_3|h|^2}{1 +\ssnr_1|h|^2}) + \sqrt{\frac{1}{(T -N)B}\left(1 - \frac{1}{(\ssnr_1|h|^2 + 1)^2}\right) } Q^{-1}(\e)\log_2e}{\sqrt{\frac{1}{(T -N)B}\left(1 - \frac{1}{(\ssnr_3|h|^2 + 1)^2}\right) }\log_2e}\right).
\end{align}
\hrule
\end{figure*}
Note that, the error probability $\e'_{|h|^2}$ again varies with the fading coefficient $|h|$.
\item In scenario 4, the constant error probability $\e$ is attained with rate $r_2(\ssnr_4, |h|^2)$.
\end{itemize}
By combining the above error probability expressions, the average probability of error for variable-rate transmissions is given by
\begin{equation}
\label{eq:avg_error}
\begin{split}
\e_{avg} = &\Pr(\mH_{1},\hH_{1})\e + \Pr(\mH_{1},\hH_{0})\mathbb{E}_{\vert h \vert^2}\{\e''_{|h|^2}\} \\ &+\Pr(\mH_{0},\hH_{1})\mathbb{E}_{\vert h \vert^2}\{\e'_{|h|^2}\} + \Pr(\mH_{0},\hH_{0})\e.
\end{split}
\end{equation}
We can further express $\e_{avg}$ by using the prior probabilities of the channel state given in (\ref{eq:prior_probs}) and the probabilities of channel sensing decisions in (\ref{false_alarm_probability}) -- (\ref{eq:sensing_condprobs}) as follows:
\begin{equation}
\begin{split}
\e_{avg} &= \Pr(\mH_{1})\Pr(\hH_{1}|\mH_{1})\e + \Pr(\mH_{1})\Pr(\hH_{0}|\mH_{1})\mathbb{E}_{\vert h \vert^2}\{\e''_{|h|^2}\} \\ &\hspace{0.4 cm}+ \Pr(\mH_{0})\Pr(\hH_{1}|\mH_{0})\mathbb{E}_{\vert h \vert^2}\{\e'_{|h|^2}\} + \Pr(\mH_{0})\Pr(\hH_{0}|\mH_{0})\e \\
&=\frac{q}{q+s}P_d\e + \frac{q}{q+s}(1 - P_d)\mathbb{E}_{\vert h \vert^2}\{\e''_{|h|^2}\} \\&\hspace{0.4cm}+ \frac{s}{s+q}P_f\mathbb{E}_{\vert h \vert^2}\{\e'_{|h|^2}\} + \frac{s}{s+q}(1 - P_f)\e.
\end{split}
\end{equation}
Now we can obtain the transition probabilities in a similar fashion as in Section \ref{sec:state-transition} for $i = 1,2,3,4$ and $k = 5,6,7,8$:
\begin{equation}
\small
\begin{split}
\label{eq:probs_perfectCSI}
\begin{array}{ll}
p_{i1} = (1 - s)P_d(1 - \e) & p_{k1} = qP_d(1 - \e),\\
p_{i2} = (1 - s)P_d\e & p_{k2} = qP_d\e,\\
p_{i3} = (1 - s)(1 - P_d)(1-\e''_{|h|^2})& p_{k3} = q(1 - P_d)(1-\e''_{|h|^2}),\\
p_{i4} = (1 - s)(1 - P_d)\e''_{|h|^2} & p_{k4} = q(1 - P_d)\e''_{|h|^2},\\
p_{i5} = sP_f(1-\e'_{|h|^2}) & p_{k5} = (1 - q)P_f(1-\e'_{|h|^2}),\\
p_{i6} = sP_f\e'_{|h|^2} & p_{k6} = (1 - q)P_f\e'_{|h|^2},\\
p_{i7} = s(1 - P_f)(1-\e) & p_{k7} = (1 - q)(1 - P_f)(1-\e),\\
p_{i8} = s(1 - P_f)\e & p_{k8} = (1 - q)(1 - P_f)\e,
\end{array}
\end{split}
\normalsize
\end{equation}
where the transition probabilities to states 1, 2, 7 and 8 are constant while the rest of the transition probabilities depend on the fading coefficient $|h|$.
\subsubsection{Throughput Under Buffer Limitations}
We will use the same techniques described in Section \ref{sec:throughput_underQoS_noCSI}. Since service rates in ON states are functions of the fading coefficient in variable-rate transmission, the only difference comes from the moment generating functions of the processes in ON states as follows:
\begin{equation}
\begin{split}
\label{eq:diagonal_matrix_perfectCSI}
\phi(\theta) = \text{diag}\Big\{&\mathbb{E}_{|h|^2}\{e^{\theta (T - N)Br_1(\ssnr_1, |h|^2)}\} ,1,\\ &\mathbb{E}_{|h|^2}\{e^{-\theta (T - N)Br_2(\ssnr_4, |h|^2)}\},
1,\\
&\mathbb{E}_{|h|^2}\{e^{\theta (T - N)Br_1(\ssnr_1, |h|^2)}\},1, \\ &\mathbb{E}_{|h|^2}\{e^{-\theta (T - N)Br_2(\ssnr_4, |h|^2)}\},1\Big\}.
\end{split}
\end{equation}
Then, the approach given in Section \ref{sec:throughput_underQoS_noCSI} can be applied to obtain the effective rate under QoS constraints as in (\ref{eq:eff-cap-cognitive_perfectCSI}) on the next page. The target error probability $\e$ can be optimized to maximize the effective throughput.
When the cognitive radio channel is not subject to any buffer constraints, hence QoS exponent $\theta \to 0$, we have the effective rate expression given in (\ref{eq:eff-cap-theta-zero_perfectCSI}).

\begin{figure*}
\begin{equation}
\label{eq:eff-cap-cognitive_perfectCSI}
\small
\begin{split}
R_E(\tSNR, \theta) = &\max_{\e\geq0}-\frac{1}{\theta TB}\log_e\mathbb{E}_{|h|^2}\Bigg(\frac{1}{2}\Big[(p_{i1} + p_{k5})\mathbb{E}_{|h|^2}\{e^{-\theta (T - N)Br_1(\ssnr_1, |h|^2)}\} + (p_{i3} + p_{k7})\mathbb{E}_{|h|^2}\{e^{-\theta (T - N)Br_2(\ssnr_4, |h|^2)}\} \\+ &p_{i2} + p_{i4}+ p_{k6} + p_{k8}\Big] + \frac{1}{2}\Big\{\Big[(p_{i1} - p_{k5})\mathbb{E}_{|h|^2}\{e^{-\theta (T - N)Br_1(\ssnr_1, |h|^2)}\} + (p_{i3} - p_{k7})\mathbb{E}_{|h|^2}\{e^{-\theta (T - N)Br_2(\ssnr_4, |h|^2)}\} \\+ &p_{i2} + p_{i4}- p_{k6} - p_{k8}\Big]^2 + 4(p_{k1}\mathbb{E}_{|h|^2}\{e^{-\theta (T - N)Br_1(\ssnr_1, |h|^2)}\} + p_{k2} + p_{k3}\mathbb{E}_{|h|^2}\{e^{-\theta (T - N)Br_2(\ssnr_4, |h|^2)}\} + p_{k4})\\ &\times(p_{i5}\mathbb{E}_{|h|^2}\{e^{-\theta (T - N)Br_1(\ssnr_1, |h|^2)}\} + p_{i6} + p_{i7}\mathbb{E}_{|h|^2}\{e^{-\theta (T - N)Br_2(\ssnr_4, |h|^2)}\} + p_{i8})\Big\}^{\frac{1}{2}}\Bigg)
\end{split}
\normalsize
\end{equation}
\hrule
\end{figure*}

\begin{figure*}
\begin{equation}
\label{eq:eff-cap-theta-zero_perfectCSI}
\begin{split}
R_E(\tSNR, 0) = \max_{\e\geq0}&\frac{(T - N)\mathbb{E}_{\vert h \vert^2}\{r_1(\ssnr_1, |h|^2)\}P_d}{2T(s + q)}\Big((1 - s)(3q - s) + 4sq\Big)(1 - \e) \\&+ \frac{(T - N)\mathbb{E}_{\vert h \vert^2}\{r_2(\ssnr_4, |h|^2)\}(1 - P_d)}{2T(s + q)}\Big((1 - s)(3q - s) + 4sq\Big)(1 - \mathbb{E}_{\vert h \vert^2}\{\e''_{|h|^2}\})\\&+\frac{(T - N)\mathbb{E}_{\vert h \vert^2}\{r_1(\ssnr_1, |h|^2)\}P_f}{2T(s + q)}\Big((1 - s)(3s - q) + 4sq\Big)(1 - \mathbb{E}_{\vert h \vert^2}\{\e'_{|h|^2}\})\\&+ \frac{(T - N)\mathbb{E}_{\vert h \vert^2}\{r_2(\ssnr_4, |h|^2)\}(1 - P_f)}{2T(s + q)}\Big((1 - s)(3s - q) + 4sq\Big)(1 - \e)
\end{split}
\end{equation}
\hrule
\end{figure*}

\section{Numerical Results}
\label{sec:sim_results}
In this section, the results of numerical computations are illustrated. More specifically, we numerically investigate optimal transmission parameters such as optimal fixed transmission rates and optimal target error probabilities in variable-rate transmissions. Furthermore, we analyze the impact sensing parameters and performance (e.g., sensing duration and threshold, and detection and false-alarm probabilities), different levels of QoS constraints, and codeword blocklengths on the throughput in cognitive radio systems. Numerically, we provide characterizations for key tradeoffs.

In the simulations, we consider Rayleigh fading channel with exponentially distributed fading power with unit mean, i.e., $f_{|h|^2} (|h|^2) = e^{-|h|^2}$. It is assumed that the channel bandwidth $B = 10$ kHz, noise power $\sigma_n^2 = 0.05$, interference power $\sigma_s^2 = 0.12$ and $\mathbb{E}\{|g_{sp,j}|^2\}=1$. In the two state Markov model, the transition probabilities from busy to idle state $P_{B, I} = s$ and from idle to busy state $P_{I, B} = q$ are set to $0.6$ and $0.2$, respectively. The average power values are $\overline{P}_1 = 0$ dB and $\overline{P}_2 = 10$ dB in the cases of channel being sensed to be busy and idle, respectively. Sensing threshold $\lambda$ is chosen as $0.1$ in order to have reasonable probabilities of false alarm and detection. In this case, we have $P_d \approx 0.863$ and $P_f \approx 0.005$. Unless mentioned explicitly, frame duration $T$ is $100$ ms, sensing duration $N$ is $1$ ms, and hence data transmission is performed with $(T - N)B = 990$ complex signal samples.

\subsection{Fixed-Rate Transmissions}

\begin{figure}[htb]
\centering
\includegraphics[width=\figsize\textwidth]{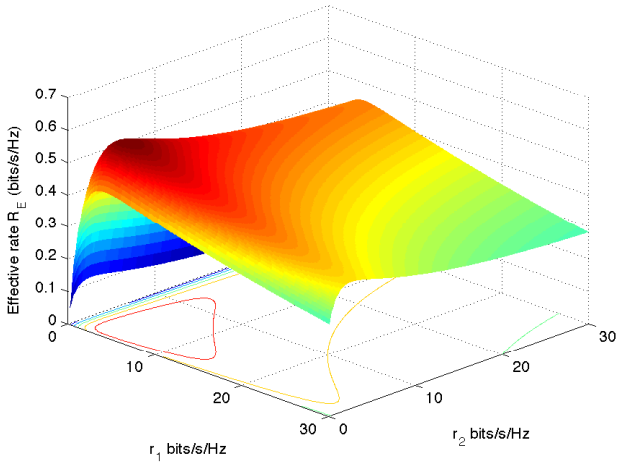}
\caption{The effective rate $R_E$ vs. fixed transmission rates $r_1$ and $r_2$ in the Rayleigh fading environment. The code blocklength is $(T-N)B = 990$.}
\label{fig:RE_tranmissionrates}
\end{figure}

In Fig. \ref{fig:RE_tranmissionrates}, the effective rate $R_E$ is plotted as a function of fixed transmission rates $r_1$ and $r_2$. The QoS exponent $\theta$ is set to $0.001$.  We see that effective rate is maximized at unique $r_1$ and $r_2$ values.
\begin{figure}[htb]
\centering
\includegraphics[width=\figsize\textwidth]{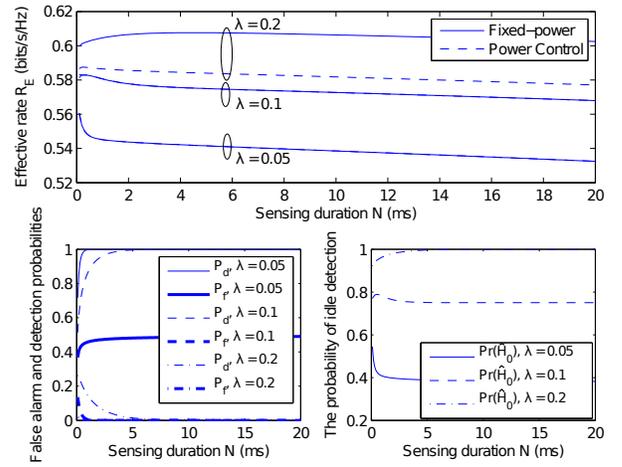}
\caption{The effective rate $R_E$, the probabilities of false alarm $P_f$ and detection $P_d$, the probability of idle detection $\Pr(\hH_{0})$ vs. sensing duration $N$ in fixed-rate transmission.}
\label{fig:effectiverate_sensing_noCSI}
\end{figure}

We analyze the tradeoff between the sensing duration $N$ and the effective rate $R_E$. Hence, in Fig. \ref{fig:effectiverate_sensing_noCSI}, we plot the effective rate, the probabilities of false alarm and detection, the probability of idle detection $\Pr(\hH_{0})$ as a function of the channel sensing duration $N$ for $\lambda = 0.05, 0.1$ and $0.2$. The QoS exponent $\theta$ is set to $0.001$. Again, fixed-rate transmissions are considered and the effective rate is maximized over transmission rates. For $\lambda = 0.05$, the false alarm and detection probabilities increase to $1$ and approximately $0.5$, respectively with increasing $N$.  Since the false alarm probability is higher, we have lower probability of detecting channel as idle as seen in the lower right figure. Hence, the channel is not efficiently utilized by cognitive users due to imperfect channel sensing decisions. Therefore, the effective rate is small. On the other hand, when $\lambda = 0.2$, we have lower false alarm and detection probabilities since the threshold level in hypothesis testing is higher. The probabilities of false alarm and detection diminish to $0$ as $N$ increases. Thus, the secondary user senses the channel as idle more frequently and performs data transmission with higher average power level, which leads to higher effective rate. But, this comes at the expense of higher interference on the primary users, which may be prohibitive since primary users' transmission cannot be sufficiently protected. If we impose the average interference power constraint in (\ref{eq:avg-interference-power}) with  $\frac{I_0}{\max_{j} \mathbb{E}\{|g_{sp,j}|^2\}}=7$ dB, and peak transmission power constraints $0$ dB and $10$ dB for $\overline{P}_1$ and $\overline{P}_2$, respectively, the power level is limited by the interference constraint for lower values of detection probability. Hence, we have lower effective rate with power control imposed through the constraint in (\ref{eq:avg-interference-power}) when $\lambda = 0.2$. As a result, we provide effective protection for primary users. In the case of $\lambda = 0.1$, reliable channel sensing is achieved since the probabilities of false alarm and detection approach $0$ and $1$, respectively. The effective rate increases until a certain threshold due to reliable channel sensing. However, after that threshold, the effective rate decreases with increasing channel sensing duration. The reason is that as channel sensing takes more time, less time is available for data transmission. Additionally, shorter coding blocklength for data transmission further affects adversely, leading to lower effective throughput. Thus, there is a more intricate tradeoff between channel sensing duration and throughput in the finite blocklength regime.

\begin{figure}[htb]
\centering
\includegraphics[width=\figsize\textwidth]{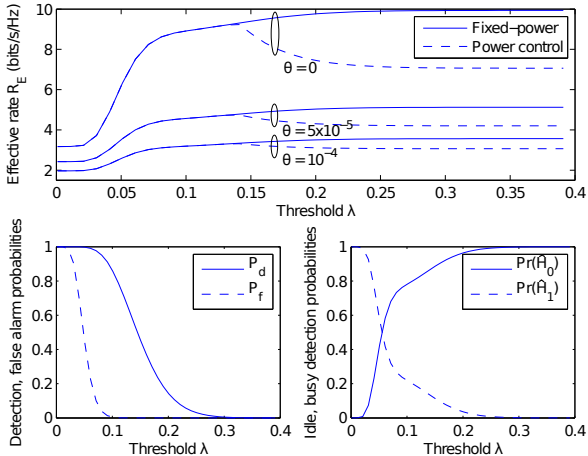}
\caption{The effective rate $R_E$, the probabilities of detection and false alarm, probabilities of idle and busy detection vs. sensing threshold $\lambda$ in fixed-rate transmissions.}
\label{fig:effectiverate_lambda_noCSI}
\end{figure}

In order to analyze the impact of the choice of the sensing threshold on the effective rate, in Fig. \ref{fig:effectiverate_lambda_noCSI}, we plot the effective rate, probabilities of false alarm and detection, probabilities of idle and busy detection vs. sensing threshold $\lambda$ for the values of QoS exponent $\theta = 0$, $5\times10^{-5}$ and $10^{-4}$ in the fixed-rate transmission case.  Since the channel sensing method is independent of $\theta$, we display the behavior of the above-mentioned probabilities without any buffer limitations in the lower subfigures. The effective rate is again maximized with respect to transmission rates. Initially, as $\lambda$ increases, the probability of false alarm starts to diminish. This improves the detection performance, and hence secondary users obtain more accurate channel sensing results. Therefore, the effective rate starts increasing. As $\lambda$ continues to increase, the false alarm probability approaches $0$ and the probability of detection starts to decrease as well.  Hence, the cognitive users fail to detect the primary users' activity even if they are active in the channel (i.e., we have higher miss detection probability), and use the channel more frequently by transmitting data with higher average power level, which explains the second increase in the effective rate. However, experiencing significant interference can deteriorate the primary users' data transmission. To avoid this harmful interference caused by the secondary user, the lower bound on the detection probability can be imposed, i.e., $P_d \ge 0.6$. Also, the transmission power $\overline{P}_2$ can be limited by the average interference constraint in (\ref{eq:avg-interference-power}) with $\frac{I_0}{\max_{j} \mathbb{E}\{|g_{sp,j}|^2\}}=7$ dB , which leads to decreasing effective rate as the secondary users fail to detect the primary users' activity. In the figure, we also see that effective rate decreases with increasing $\theta$. Thus, the effective rate takes the highest values in the absence of QoS constraints, i.e., when $\theta = 0$.

\subsection{Variable-Rate Transmissions}

\begin{figure}[htb]
\centering
\includegraphics[width=\figsize\textwidth]{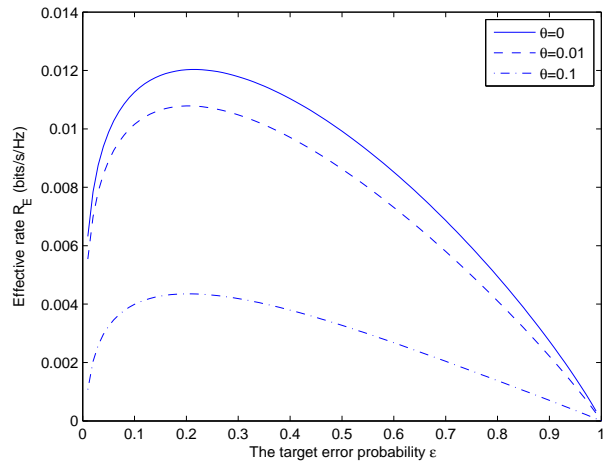}
\caption{The effective rate $R_E$ vs the probability of error $\epsilon$ for different values of QoS exponent $\theta$ in variable-rate transmission.}
\label{fig:RE_vsepsilon}
\end{figure}

In Fig. \ref{fig:RE_vsepsilon}, we consider variable-rate transmissions, and display numerical results for the effective rate as a function of the target error probability $\e$ for $\theta = 0, 0.01$ and $0.1$. As larger values of the error probability $\e$ indicate that cognitive users' data transmission is subject to more errors, they enter into OFF states frequently, where rate of reliable transmission is effectively zero.  Therefore, effective rate decreases as $\e$ increases beyond a threshold. We also observe that effective rate is maximized at a unique optimal error probability $\epsilon$. Moreover, effective rate decreases as QoS constraints become more stringent (i.e., for larger values of $\theta$).

\begin{figure}[htb]
\centering
\includegraphics[width=\figsize\textwidth]{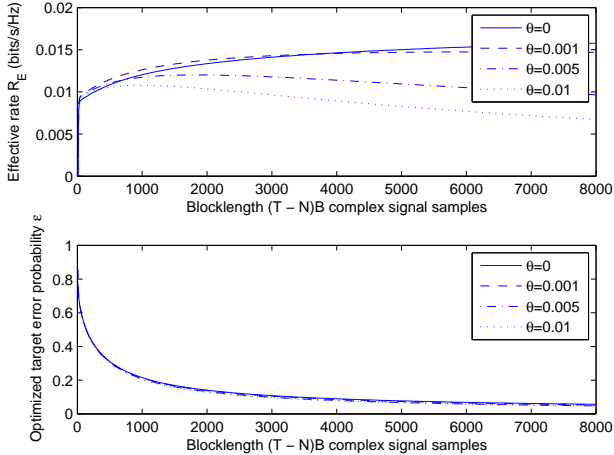}
\caption{The effective rate $R_E$ and the probability of error $\epsilon$ vs. blocklength $(T - N)B$ in variable-rate transmission.}
\label{fig:RE_epsilon_vsblocklength}
\end{figure}

The tradeoff between the blocklength and effective rate in variable-rate transmission is analyzed. Hence, in Fig. \ref{fig:RE_epsilon_vsblocklength}, we display the behavior of the optimized error probability and effective rate as a function of the code blocklength $(T-N)B$ for $\theta = 0,  0.001, 0.005, 0.01$. In the lower subfigure we see that as the code blocklength increases, the optimal error probability, which maximizes the effective rate, decreases for given $\theta$ values. In the upper subfigure, we observe that if there is no such buffer limitation, effective rate increases with increasing blocklength. However, under buffer constraints with $\theta = 0.005$ and $0.01$, as code blocklength increases until a certain threshold, data transmission is performed with decreasing error probability $\epsilon$, which improves the system performance because longer codewords are transmitted more reliably. On the other hand, the effective rate starts to decrease after the threshold. This is due to our assumption that fading stays constant over the frame of $T$ seconds. As the blocklength and hence the value of $T$ increase, cognitive users experience slower fading. Therefore, possible unfavorable deep fading lasts longer, leading to degradation in performance. In order to avoid buffer overflows, secondary transmitter becomes more conservative and supports only smaller arrival rates.

\begin{figure}[htb]
\centering
\includegraphics[width=\figsize\textwidth]{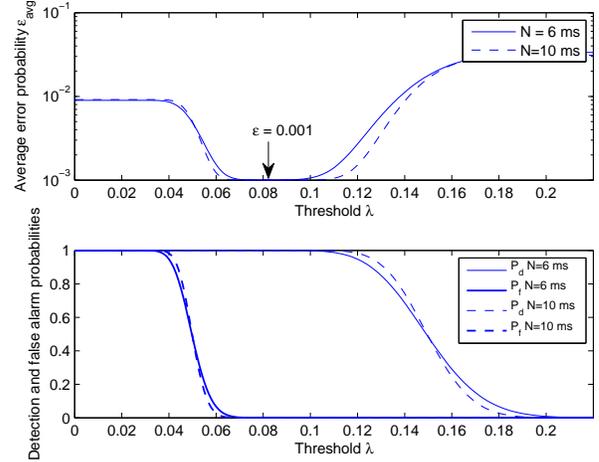}
\caption{The average error probability $\epsilon_{avg}$ and probabilities of false-alarm, detection vs. sensing threshold $\lambda$.}
\label{fig:Avgerror_vsthreshold}
\end{figure}

In Fig. \ref{fig:Avgerror_vsthreshold}, we plot the average error probability $\e_{avg}$, which maximizes the effective rate in variable-rate transmission and the probabilities of detection and false alarm vs. sensing threshold $\lambda$ for sensing duration of $N = 6$ ms and $10$ ms. In the presence of CSI knowledge at the transmitter, secondary transmitter performs variable-rate data transmission with given fixed target error probability $\e = 0.001$ and $\theta=0.001$. As we know from the analysis in Section \ref{sec:state_transition_perfectCSI}, error probability does not stay fixed at the target level of $\e$ in scenarios 2, 3 where busy channel is sensed as idle and idle channel is sensed as busy, respectively. As $\lambda$ increases, the probability of false alarm starts decreasing. Hence, average error probability decreases. When the probability of detection and the probability of false alarm approach 1 and 0, respectively (in the case of perfect channel sensing), the average error probability is equal to the fixed target error probability $\e = 0.001$. As $\lambda$ continues to increase, the detection probability diminishes and miss detection (scenario 2) occurs more frequently, resulting in error probabilities greater than $\e$. Cognitive users can experience frequent errors in miss detections with variable error probability $\e'_{|h|^2}$, which is larger than the fixed target error probability of $\e$. Therefore, we have higher average error probability. We can see that channel sensing plays a critical role on the average error probability in variable-rate transmissions. Finally we note that as sensing duration increases, the probabilities of false alarm and detection decrease with higher slopes as threshold increases. We also note that lower average error probability is achieved with larger $N$ values when $0.05 < \lambda \le 0.14$.

\begin{figure}[htb]
\centering
\includegraphics[width=\figsize\textwidth]{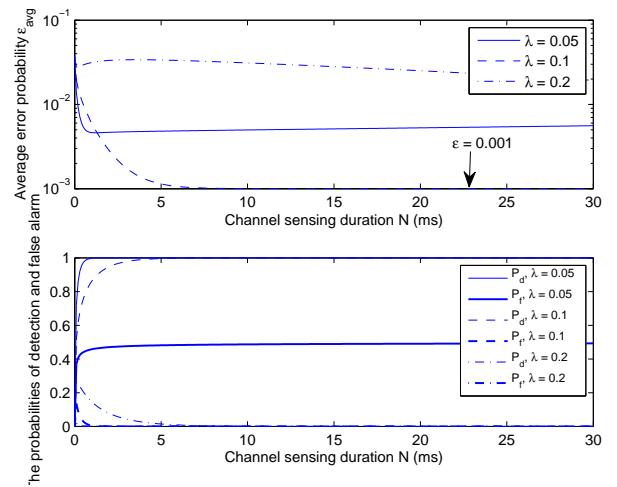}
\caption{The average error probability $\epsilon_{avg}$ and probabilities of false-alarm, detection vs. channel sensing duration $N$.}
\label{fig:Avgerror_vssensing}
\end{figure}

Next, we analyze the tradeoff between the reliability of the variable-rate transmission and the sensing duration. In Fig. \ref{fig:Avgerror_vssensing}, the average probability of error $\e_{avg}$, which achieves the highest effective rate, the probabilities of detection and false alarm are given as a function of sensing duration $N$ for $\lambda = 0.05, 0.1$ and $0.2$. The target error probability $\e$ is fixed to $0.001$. When $\lambda = 0.05$, the detection probability approaches $1$ and the false alarm probability approaches $0.5$ as sensing duration increases. Thus, cognitive users detect the channel as busy more and transmit data with fixed error probability $\e$ or variable error probability $\e'_{|h|^2}$ (scenario 1 and scenario 3, respectively).  The average error probability decreases when channel sensing takes more time and approaches approximately $\e$. For $\lambda = 0.1$, cognitive users almost perfectly sense the channel with false alarm and detection probabilities approaching $0$ and $1$, respectively with increasing sensing duration.  Thus, average error probability decreases and approaches $\e = 0.001$.  Therefore, data transmission is performed at the target error rate. If $\lambda$ is chosen as $0.2$, error probability increases until a certain threshold since we have lower false alarm and detection probabilities and the channel is detected as idle even though it is occupied by primary users, where cognitive users' transmission rate is achieved with error rate $\e''_{|h|^2}$ that is much bigger than the target error probability $\e$. After that threshold, less time is allocated for data transmission. Therefore, lower transmission rates are supported, yielding more reliable data transmission, and hence decreasing the average error probability.

\subsection{Fixed-Rate vs. Variable-Rate Transmissions}
In this subsection, we compare the effective rate achieved under fixed-rate and variable-rate transmission schemes.
\begin{figure}[htb]
\centering
\includegraphics[width=\figsize\textwidth]{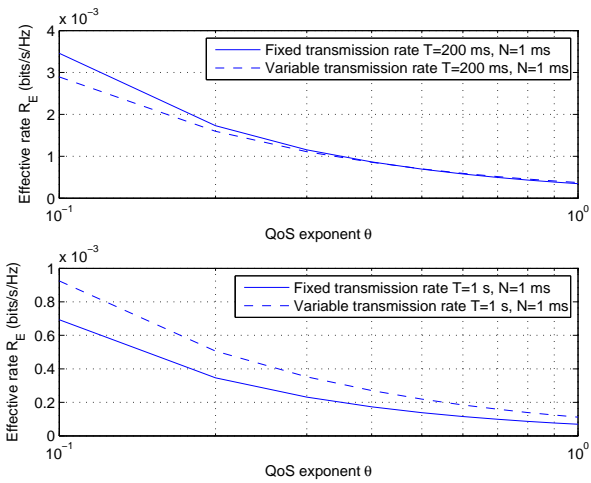}
\caption{The effective rate $R_E$ vs. QoS exponent $\theta$ for fixed-rate and variable-rate transmission for different $T$ values.}
\label{fig:RE_vstheta}
\end{figure}

In Fig. \ref{fig:RE_vstheta}, we display numerical results for the effective rate vs. QoS exponent $\theta$ in fixed-rate and variable-rate transmissions for $T=200$ ms, $N=1$ ms and $T=1$ s, $N=1$ ms. Larger values of $\theta$ indicate that data transmission is performed under more strict QoS constraints. We see that increasing $\theta$ diminishes the effective rate $R_E$ for both transmission schemes. The variable-rate transmission achieves better performance when $T=1$ s, $N=1$ ms for all values of $\theta$. On the other hand, fixed-rate transmission outperforms for low values of $\theta$ when $T=200$ ms, $N=1$ ms. Under more strict buffer limitations (higher values of $\theta$), cognitive users send data with lower rates. Thus, the reliability of transmission becomes more important. Therefore, instead of sending data at constant rates, transmitter benefits more by varying the rate.

\begin{figure}[htb]
\centering
\includegraphics[width=\figsize\textwidth]{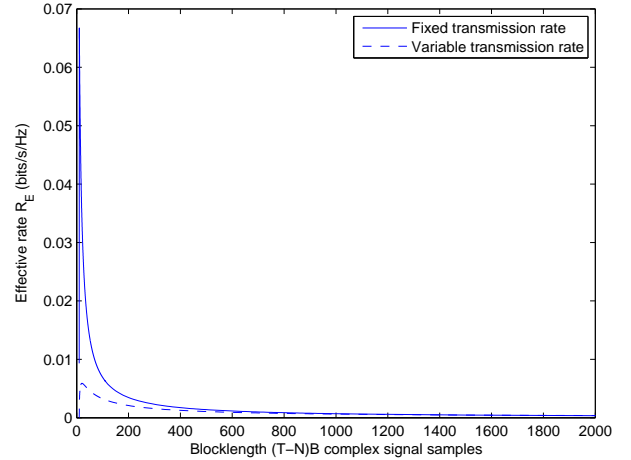}
\caption{The effective rate $R_E$ vs. blocklength $(T - N)B$ for fixed-rate and variable-rate transmission, $\theta=1$.}
\label{fig:RE_vsblocklength}
\end{figure}

Effective rate $R_E$ is given as a function of blocklength $(T - N)B$ for fixed-rate and variable-rate transmissions in Fig. \ref{fig:RE_vsblocklength}. We previously observed that effective rate increases until a certain threshold with increasing code blocklength. After that threshold, effective rate starts to diminish. The reason of this trend is explained in Fig. \ref{fig:RE_epsilon_vsblocklength} for variable-rate transmission. In this figure, we also see that the same behavior is observed for fixed-rate transmission. We interestingly note that transmitting with constant rates leads to higher effective rate compared to varying the rate based on channel conditions when code blocklength is less than $1500$ complex signal samples. When $(T - N)B$ is increased beyond $1500$ complex signal samples, keeping the error probability constant and performing data transmission with variable rate result in better performance.

\section{Conclusion}
\label{sec:conclusion}
In this paper, we have analyzed the throughput of cognitive radio systems in the finite blocklength regime under buffer constraints.  Through the effective capacity formulation, we have characterized the maximum constant arrival rates that the cognitive radio channel can support with finite blocklength codes while satisfying statistical QoS constraints imposed as limitations on the buffer violation probability. We have first focused on the scenario in which the CSI of the secondary link is assumed to be perfectly known at the secondary receiver only. In this case, the secondary transmitter sends the data at two different constant rate levels, which depend on the channel sensing decision, and error rates vary with the channel conditions. In the second scenario, perfect CSI is available at both the secondary transmitter and receiver. Under this assumption, the secondary transmitter, considering a target error rate level, varies its transmission rate according to the time-varying channel conditions. For both scenarios, we have determined the throughput as a function of state transition probabilities of the cognitive radio channel, prior probabilities of idle/busy state of primary users, sensing decisions and reliability, the block error probability, QoS exponent, frame and sensing durations.


We have investigated the interactions and tradeoffs between different buffer, sensing, transmission, and channel parameters and the throughput.
Through the numerical results, we have demonstrated that sensing threshold, duration and reliability have significant impact on the performance. In particular, we have observed that highly inaccurate sensing can either lead to inefficient use of resources and low throughput or cause possibly high interference on the primary users. We have also noted that sensing-throughput tradeoff is more involved since increasing the sensing duration for improved sensing performance not only decreases the time allocated to data transmission but also results in shorter codewords being sent, lowering the transmission reliability. Additionally, we have seen in the case of variable transmission-rate that average error probability can deviate significantly from the target error rate due to imperfect sensing. Moreover, we have remarked that throughput generally decreases as the QoS exponent $\theta$ increases (i.e., as QoS constraints become more stringent), and variable-rate transmissions have better performance under more strict QoS restrictions while fixed-rate transmissions lead to higher throughput under looser QoS constraints.


\vspace{-1cm}
\begin{IEEEbiographynophoto}{Gozde Ozcan}
received the B.S. degree in Electrical and Electronics Engineering from Bilkent University, Ankara, Turkey in 2011. She is currently working towards the Ph.D. degree in the Department of Electrical Engineering and Computer Science, Syracuse University. Her research interests are in the fields of wireless communications, statistical signal processing and information theory. Currently, she has particular interest in cognitive radio systems.
\end{IEEEbiographynophoto}

\begin{IEEEbiographynophoto}{Mustafa Cenk Gursoy}
 received the Ph.D. degree in electrical engineering from Princeton University, Princeton, NJ, in 2004, and the B.S. degree in electrical and electronics engineering from Bogazici University, Istanbul, Turkey, in 1999 with high distinction. He was a recipient of the Gordon Wu Graduate Fellowship from Princeton University between 1999 and 2003. In the summer of 2000, he worked at Lucent Technologies, Holmdel, NJ, where he conducted performance analysis of DSL modems. Between 2004 and 2011, he was a faculty member in the Department of Electrical Engineering at the University of Nebraska-Lincoln (UNL). He is currently an Associate Professor in the Department of Electrical Engineering and Computer Science at Syracuse University. His research interests are in the general areas of wireless communications, information theory, communication networks, and signal processing. He is currently a member of the editorial boards of IEEE Transactions on Wireless Communications, IEEE Transactions on Vehicular Technology, IEEE Communications Letters, and Physical Communication (Elsevier). He received an NSF CAREER Award in 2006. More recently, he received the EURASIP Journal of Wireless Communications and Networking Best Paper Award, the UNL College Distinguished Teaching Award, and the Maude Hammond Fling Faculty Research Fellowship.
\end{IEEEbiographynophoto}

\end{document}